\documentclass[iop]{emulateapj}
\slugcomment{{Accepted to ApJL:} January 15, 2014}
\usepackage{float}
\usepackage{color}

\begin{document}
\title{Forming Circumbinary Planets: N-body Simulations of Kepler-34}
\author{S. Lines$^1$, Z. M. Leinhardt$^1$, S. Paardekooper$^{2,3}$, C. Baruteau$^{2,4,5}$ and P. Thebault$^6$}
\affiliation{$^1$ School of Physics, University of Bristol, Tyndall Avenue, Bristol, BS8 1TL, UK; \emph {stefan.lines@bristol.ac.uk}\\$^2$ DAMTP, University of Cambridge, Wilberforce Road, Cambridge CB3 0WA, UK\\$^3$Astronomy Unit, School of Physics $\&$ Astronomy, Queen Mary University of London, UK\\$^4$ CNRS, IRAP, 14 avenue Edouard Belin, 31400 Toulouse, France\\$^5$ Universit{\'e} de Toulouse, UPS-OMP, IRAP, Toulouse, France\\$^6$ LESIA-Observatoire de Paris, UPMC Univ. Paris 06, Univ.~Paris-Diderot, France}

\begin{abstract}
\centering
Observations of circumbinary planets orbiting very close to the central stars have shown that planet formation may occur in a very hostile environment, where the gravitational pull from the binary should be very strong on the primordial protoplanetary disk. Elevated impact velocities and orbit crossings from eccentricity oscillations are the primary contributors towards high energy, potentially destructive collisions that inhibit the growth of aspiring planets. In this work, we conduct high resolution, inter-particle gravity enabled {\emph N}-body simulations to investigate the feasibility of planetesimal growth in the Kepler-34 system. We improve upon previous work by including planetesimal disk self-gravity and an extensive collision model to accurately handle inter-planetesimal interactions. We find that super-catastrophic erosion events are the dominant mechanism up to and including the orbital radius of Kepler-34(AB)b, making {\it in-situ} growth unlikely. 
It is more plausible that Kepler-34(AB)b migrated from a region beyond 1.5 AU. Based on the conclusions that we have made for Kepler-34 it seems likely that all of the currently known circumbinary planets have also migrated significantly from their formation location with the possible exception of Kepler-47(AB)c.
\end{abstract}

\section{Introduction}

\begin{figure*}[tbph]
\centering
\includegraphics[scale=0.45]{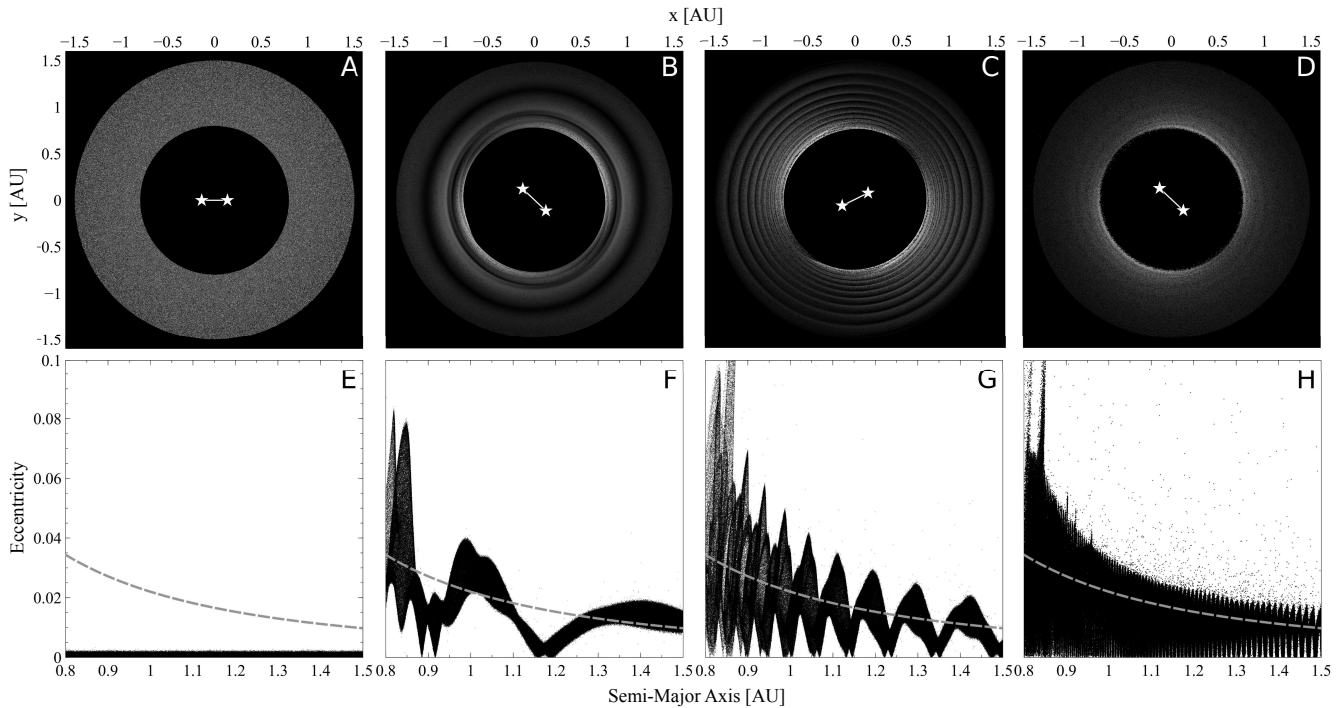}
\caption{Evolution of eccentricity waves at $t_0$ ({\bf A}), $25\,P_{AB}$ ({\bf B}), $100\,P_{AB}$ ({\bf C}), $1000\,P_{AB}$ ({\bf D}). The instantaneous eccentricity of each planetesimal is shown with a grey value between black ($e=0$) and white ($e=0.02$ {\bf A}, $e=0.1$ {\bf B-D}). {\bf E}-{\bf H} show eccentricity evolution versus $a$ with $e_{ff}$ \citep{paardekooper12} shown in grey.}
\label{fig:multi}
\vspace{+10pt}
\end{figure*}

Several planets have been discovered by Kepler in `p-type' orbits fully encompassing tight binaries \citep{doyle11, welsh12, orosz12, schwamb13}. These planets orbit close to their host binaries ($a < 1.1$ AU) and are subject to significant perturbations which cause impact speeds to increase making accretion close to the binary difficult \citep{marzari00,thebault06,scholl07}. 

\cite{paardekooper12} conducted numerical simulations of test particles within circumbinary planetesimal disks including a collision model from \citet{stewart09} which allowed for accretion and erosion. The collision model provided an alternative mechanism for planetesimal growth: accretion of dust created in catastrophic collisions. However, the authors found that growth at the location of the Kepler-16 and Kepler-34 planets was difficult. Their investigation also revealed the contribution of short-period perturbations on the disk. This fast eccentricity forcing evolves more quickly than the gas damping timescale therefore the disk stays excited.

Thus, there remains a degree of uncertainty about whether observed circumbinary planets formed in-situ. In addition, the literature does not agree on the locations that could have supported the planets' growth within a circumbinary disk. For example, \cite{meschiari12a} identify a narrow range of annuli just outside of 1 AU that could sustain planetesimal growth in Kepler-16 and \cite{paardekooper12} find the accretion friendly zone is beyond 4 AU.

Most previous work omit inter-planetesimal gravity and/or a comprehensive collision model. In this paper, we present high resolution, 3D, inter-particle gravity (IPG) enabled {\emph{N}}-body simulations of a circumbinary protoplanetary disk in order to address the question of where the circumbinary planets can form. We focus on the orbital dynamics, collisional evolution and physical growth of 100 kilometre-sized planetesimals in the Kepler-34 system. We consider a purely $N$-body case to isolate the effects of inter-planetsimal gravity, leaving the inclusion of a gas disk for future work.

To account for oblique, high speed impacts from orbit crossings, we use a state-of-the-art collision model \citep[][Leinhardt et al.~in prep]{leinhardt12}. This collision model more accurately identifies regions where planetesimal growth can occur than previous work. We use statistical arguments to classify the feasibility of sustained growth events in the disk, addressing the primary question of whether in-situ growth of Kepler-34(AB)b is possible. 

In section \ref{sec:theory} we discuss the analytics of circumbinary planetesimals acting under perturbations from the stellar binary. Our numerical method and collision model are outlined in section \ref{sec:method}. In section \ref{sec:results} we present our results and section \ref{sec:dis} discusses the broader implications. Conclusions are drawn in section \ref{sec:summary}.

\section{Perturbations on the planetesimal disk}\label{sec:theory}

The motion of a planetesimal in a circumbinary disk can be approximated by a Keplerian orbit about the primary star plus perturbations by the secondary. These perturbations are described by a disturbing function which accounts for the additional acceleration caused by the perturbing potential. The secular effects can be determined by averaging the disturbing function over the binary and planetesimal orbital motion. The overall dynamical effect is the oscillation of planetesimal eccentricities about the forced eccentricity \citep{moriwaki04},
\begin{equation}
e_f =  \frac{5}{4}\frac{M_A-M_B}{M_*}\frac{a_b}{a}e_b\frac{1+\frac{3e_b^2}{4}}{1+\frac{3e_b^2}{2}},
\end{equation}
where $M_A$ and $M_B$ are the individual star masses, $M_* = M_A + M_B$ and $a_b$, $e_b$, $a$ and $e$ are the binary and planetesimal semi-major axis and eccentricity respectively. The spatial frequency of these oscillations increases with time until orbit crossings of planetesimals on low and high eccentricity orbits occur.

Planetesimals on circular orbits around the point-mass potential of the binary are subject to short-period eccentricity oscillations on the orbital timescale of the planetesimals, characterised by a forced eccentricity $e_{ff}$ which falls off as $a^{-2}$ (see Figure 1) \citep{moriwaki04,paardekooper12}. \cite{rafikov13} suggested that for a non-eccentric binary adjusting the initial azimuthal velocity of the planetesimals could remove these oscillations. However, we find that this adjustment has only a minor impact on the eccentricities and collision speeds.

For circumbinary systems with $M_A$ = $M_B$ or $e_b$ = 0, secular effects are absent, since $e_f \longrightarrow 0$ for these special cases, so only the fast term remains. By investigating the equal mass Kepler-34 system, the influence of the fast eccentricity forcing can be isolated. 

\section{Numerical Methods}\label{sec:method}

We simulate planetesimal and binary interactions using the parellelised gravitational {\emph{N}}-body code {\emph{PKDGRAV}} \citep{richardson00,stadel01}. The code uses a symplectic, $2^{nd}$ order, leapfrog integrator and a hierarchical tree partitioning procedure to handle the orbital integration of particles. As our simulations include computationally expensive inter-planetesimal gravity calculations, we limit the resolution of our disk to $N$=$10^6$ to allow for a confident statistical conclusion to be reached in a practical time frame. Our simulations were performed using the University of Bristol's Advanced Computing Research Centre\footnote{https://www.acrc.bris.ac.uk/}. 

\subsection{Initial Conditions}

In this paper we present three simulations; two of a circumbinary disk representative of the Kepler-34 system, and one of a control simulation around a single star. We evolve the circumbinary protoplanetary disks for 2,000 binary orbits ($\approx 150$ $yr$). The second of these simulations features identical parameters to the first, but employs a reduced IPG (RIPG) model. This is achieved by reducing the masses of planetesimals by a factor of 1000 in order to determine the effect of inter-planetesimal gravity on the collision outcome. The comparative single star simulation ($M_{\star} = 1 M_{\odot}$) uses the same timestep and disk parameters to set a benchmark for conditions known to sustain planetesimal accretion. All simulations initially contain a unimodal mass population of planetesimals in a disk of total mass $M_{Disk} = 2.8 M_{\earth}$. Our planetesimal disk masses are small in comparison to typical gas disks, and hence the effect of disk self-gravity on the dynamical evolution of planetesimals which causes orbital alignment should not be reproduced. Planetesimal bulk density, $\rho_{bulk}$ , is 2 g cm$^{-3}$, consistent with silicon-rich rocky-bodies, giving an initial planetesimal radius of $R_{pl} = 120$ km. $\rho_{bulk}$ is scaled in the RIPG case to keep $R_{pl} = 120$ km. While such large planetesimals are a far cry away from the typical 1-10 km bodies modelled in previous studies, the computational constraints on a self-gravitating planetesimal disk limit our resolution. More massive planetesimals will be harder to disrupt due to an increased binding energy, so any erosion in this case will be amplified in simulations which consider smaller bodies \citep{leinhardt12}. As such, this aspect of the model can be considered a best-case scenario.

\begin{figure}[h!]
\centering
\includegraphics[scale=0.55]{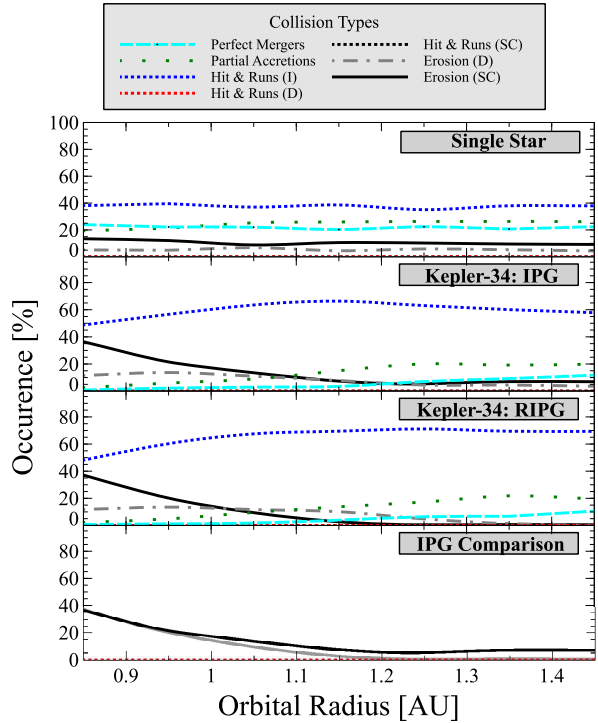}
\caption{Radial dependence of collision type (5,600 $P_{AB}$ for single star, 2,000 $P_{AB}$ for Kepler-34). Hit-and-Runs are split into projectile intact (I), disrupted (D) and supercatastrophically disrupted (SC). Erosive collisions are divided into super-catastrophic (SC, the largest post-collision remnant $<10\%$ of the colliding mass) and disruptive (D, all other erosive collisions). The RIPG collisions are identified in the comparison plot as lighter curves.}
\label{fig:collspatevo}
\vspace{+0pt}
\end{figure}

The inner ($a_i = 0.8$ AU) and outer ($a_o = 1.5$ AU) disk boundary include the current location of Kepler-34(AB)b (a $\approx$ 1.1 AU). The initial planetesimal disk has a surface density distribution  $\sigma (r) \propto r^{-1.5}$ ($\bar{\sigma} =14.8$ g cm$^{-2}$). The initial conditions of the planetesimal disk were chosen to be similar to those used in previous $N$-body simulations \citep[for example,][]{kokubo02,leinhardt05,leinhardt09}\footnote{In our simulations the expansion parameter, f, is unity.}.

The binary itself, Kepler-34(AB), consists of two stars ($M_A$ = 1.05 $M_{\odot}$ and $M_B$ = 1.02 $M_{\odot}$) each represented in the simulation by a $N$-body particle. The binary system has a semi-major axis of 0.23 AU, is highly eccentric, $e_B$ = 0.52, and has a period of 27.8 days. Due to the short period a small timestep is required (0.0025 yr) to accurately resolve the binary and maintain stability over thousands of binary orbits.

We begin the simulations, both circumbinary and single star, with an unperturbed planetesimal disk. The planetesimal inclinations and eccentricities are drawn from a Rayleigh distribution with dispersions of $\langle e^2 \rangle = 2 \langle i^2 \rangle = 0.007$ \citep{ida92}. Physical collisions are disabled during the first 1,000 binary orbits to allow the planetesimals to adopt the eccentricity and inclination dispersions imparted by the binary. 

The initial eccentricity evolution of planetesimals in the circumbinary disk is shown in Figure \ref{fig:multi}. The binary quickly perturbs the disk from the initial eccentricity distribution (Fig.~\ref{fig:multi}{\bf A} \& {\bf E}) to one with a noticeable eccentricity wave structure which is clearly visible by 25 binary orbits ($25\,P_{AB}$) (Fig.~\ref{fig:multi}{\bf B} \& {\bf F}). At $100\,P_{AB}$ (Fig.~\ref{fig:multi}{\bf C} \& {\bf G}) the frequency of the eccentricity oscillations has increased and inner disk planetesimals have reached their simulation maximum of $e \approx 0.1$. By $1000\,P_{AB}$ the planetesimals have reached a quasi-steady state with low and high eccentricity planetesimals on crossing orbits (Fig.~\ref{fig:multi}{\bf D} \& {\bf H}). This is observable in the eccentricity map as an absence of ring structures and in the $e-a$ plot by the absence of distinct peaks and troughs. The final distribution follows the derived forced eccentricity value of $e_{ff} = 0.02/a^2$ (the grey dashed line in Fig.~\ref{fig:multi}). It is at this point that we turn on collisions and allow the planetesimals to collisionally evolve.

\subsection{Collision Model}

In these simulations we use the collision model $EDACM$ which is based on \citet{leinhardt12} and has been integrated into PKDGRAV (Leinhardt et al.~in prep). EDACM is capable of handling multiple collision outcome regimes including: perfect merging, partial accretion, hit-and-run \citep[a bouncing-like effect during oblique impacts which may or may not erode the smaller projectile,][]{asphaug06,Stewart12}, partial erosion and erosive disruption. The outcome is determined using a series of scaling laws which require only the collider impact velocity, impact parameter, mass ratio and two fixed material property parameters.

\begin{figure}[h!]
\centering
\includegraphics[scale=0.42]{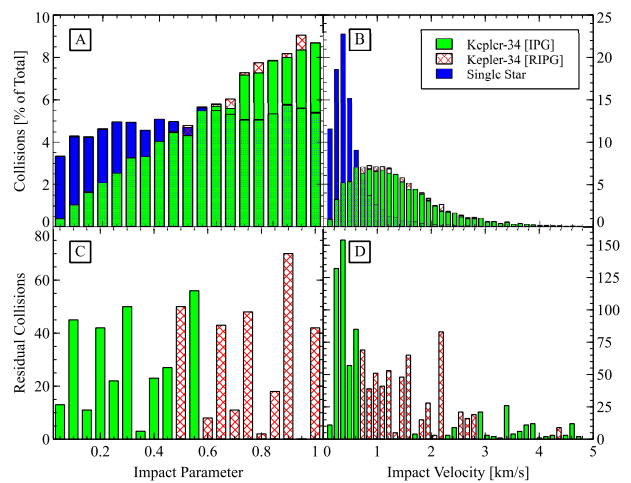}
\caption{Distribution of impact parameter (A) and impact velocity (B) for each collision for the single star (blue) and circumbinary disk. The lower panels highlight the residual collision numbers between IPG (green) and RIPG (hashed).}
\label{fig:ip}
\label{fig:iv}
\vspace{+0pt}
\end{figure}

The analytic determination of the outcome provides substantial improvements in computational efficiency and outcome accuracy over previous models such as $RUBBLE$ \citep{leinhardt05,leinhardt09} which directly integrates the planetesimal collisions by assuming that all planetesimals are gravitational aggregates.

In order to keep $N$ practical, a mass resolution limit is applied such that only particles with masses larger than their initial value, $m_0 = 1.7\times10^{21}$ g, are directly resolved. Collisions that result in the production of fragments with masses below this threshold are put into one of 10 radial bins dependent on the colliders location. The unresolved debris can be accreted by planetesimals traversing the bins. Momentum is conserved in this accretion process, however, there is no additional dynamical friction on the resolved planetesimals from the unresolved debris \citep{leinhardt05}.

\section{Results}\label{sec:results}

Figure \ref{fig:collspatevo} shows the occurrence rate of each collision type as a function of radial distance from the central star(s). In the single star case, the collision occurrence rates are almost independent of semi-major axis due to the small radial range and values of planetesimal eccentricities. In general, planetesimals around a single star grow. Growth enabling events account for $48\%$ of all collisions, while partial and catastrophic erosion events occur in $14\%$ of collisions. The remaining collisions cover hit-and-run events which are predominantly in the intact regime.

In contrast, the distribution of collision outcomes in the simulated Kepler-34 system varies radially with, excluding non-disruptive hit-and-runs, super-catastrophic collisions dominating until 1.1 AU. 

Beyond this orbital radius, disruptive collisions are exchanged for an increase in both partial and perfect mergers. Growth events contribute between 30-35$\%$ of collisions at the outer disk boundary, and only 12$\%$ at the location of Kepler-34b. In the outer regions of the simulation domain the collision occurrence rates begin converging towards that of the single star, however, the occurrence of hit-and-runs is 20$\%$ higher due to the number of oblique collisions from orbit crossings. 

The RIPG run has a 12$\%$ increase in hit-and-runs at the outer disk edge than the IPG run. This difference is due to a higher proportion of collisions in the RIPG case having a large impact parameter. However, the evolution of collision occurrence rates between the IPG and RIPG runs are almost identical.

Figure \ref{fig:ip} compares the range of impact parameter and collision speed for all collisions in each simulation. The circumbinary runs have a large number of high impact parameter collisions, $\sim 10 \%$ have $b > 0.95$. In addition, the impact speed ($v_{imp}$) is much broader in the circumbinary cases with a modal value of 1.1 km s$^{-1}$ and a maximum value of 5 km s$^{-1}$ compared to the modal $v_{imp} = 0.4 $km s$^{-1}$ and a maximum value of 3 km s$^{-1}$ in the single star case.

The absence of gravitational focusing in the RIPG circumbinary run results in a larger percentage of collisions with a high impact parameter (Fig \ref{fig:ip}C). We also observe slightly inflated impact velocities (Fig \ref{fig:ip}D). Both these effects contribute towards a more hostile environment for sustained accretion.

\begin{figure}[h!]
\centering
\includegraphics[scale=0.42,angle=0]{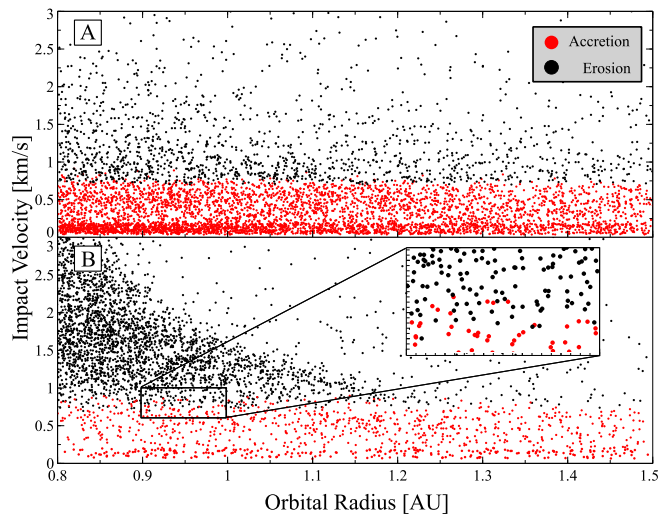}
\caption{The spatial evolution of the impact velocity for each collision in the single star case ({\bf A}) and circumbinary ({\bf B}).}
\label{collevo}
\vspace{+0pt}
\end{figure}

The radial distribution of the impact speed on collision outcome is shown in Figure \ref{collevo}. High speeds are found in the inner disk where the disk is most perturbed by the binary. A similar, albeit much weaker, effect is seen in the single star disk due to the increasing orbital velocity closer to the star. Transitions between collision regimes are not abrupt, with a notable overlap of growth and non-growth events occupying the same velocity space. In addition, planetesimals that have had a collision have a slightly higher eccentricity $< e^2 >$ = 0.03 than those that have not $<e^2>$ = 0.02. However, only $<1\%$ of planetesimals have been involved in a collision, hence a much longer simulation is required to probe the effects of collisional evolution on the dynamical temperature. These results emphasise the importance of using a collision model such EDACM, which does not rely on impact speed alone to define collision outcomes. 

\begin{figure}[h!]
\centering
\includegraphics[scale=0.38,angle=0]{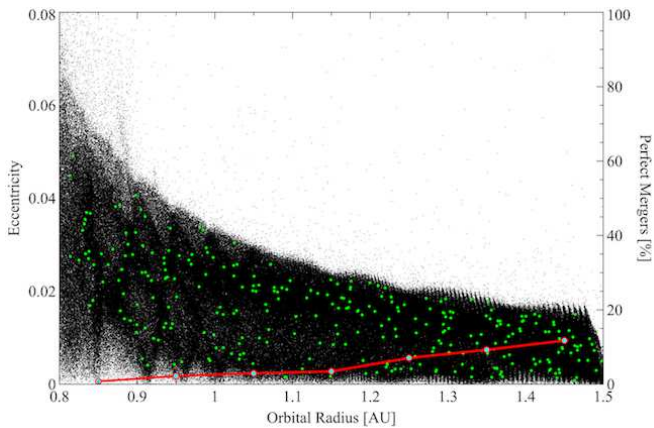}
\caption{Eccentricity distribution of planetesimals at 2,000 $P_{AB}$. Green points: $m \geq 2m_o$, black points: $m<2m_o$. The red line shows contribution from perfect mergers.}
\label{fig:growth}
\vspace{+5pt}
\end{figure}

During 2,000 $P_{AB}$ the circumbinary planetesimal disk has amassed over 10,000 collisions, yet this is a fraction of the time needed to grow planetary embryos \citep[$\sim 10^6$ yrs for a single star][]{leinhardt05}. The high proportion of non-growth events in the circumbinary case contributes to a very slow physical growth rate. Figure \ref{fig:growth} highlights planetesimals that have grown beyond $2 m_0$ (green dots). The majority of these planetesimals have increased in mass solely by perfect merging; a mechanism which contributes a maximum of only 12$\%$ at 1.5 AU and a minimum of $<1\%$ at the inner edge of the planetesimal disk. Only $0.16\%$ of the disk mass belongs to planetesimals which have grown, the equivalent of 1600 planetesimal at their initial masses. Of these, only 500 have exceeded twice their initial mass, meaning the majority of growth has likely been the result of partial accretion events, although those planetesimals with $m_p$ $<$ $2m_0$ could also be explained by disruptive collisions following perfect merging. The red line in Figure \ref{fig:growth} shows the occurrence of the total number of perfect mergers as a function of radial distance confirming a marginally higher growth rate in the outer disk. The lack of a trend in the location of accretion events and small number of grown material reveals the inability of sustained growth of material in such a disk. Instead, planetesimals appear to grow through `lucky' collisions that are a result of the occasional randomisation of their orbital velocity vectors that leads to low speed encounters. 

\section{Discussion}\label{sec:dis}

Kepler-34(AB)b orbits at 1.09 AU which falls in a regime where only 12$\%$ of collisions lead to growth (Fig.~\ref{fig:collspatevo}B). Although the eccentricity forcing falls off with $1/a^2$, $e_{ff}$ only drops to the mean initial eccentricity (e = 0.007) at 1.7 AU. However, it may not be essential to retrieve an identical environment to that of the single star case in order to sustain growth. By 1.5 AU mass changing collisions are split between partial erosion and partial accretion events, therefore, by 1.5 AU prolonged planetesimal growth becomes more sustainable. Thus, a likely scenario is that Kepler-34(AB)b formed beyond 1.5 AU and migrated inwards. It is worth mentioning that we retrieve this highly erosive disk that is unable to support sustained growth, despite our large-sized planetesimals. 

While the visual difference between the IPG and RIPG cases is small, the underlying effects propagate through to the collision outcomes. The lack of gravitational focusing with RIPG accounts for a higher proportion of oblique impacts that generally lead to an increase in hit-and-run events. While the majority of these collision are non-erosive, they do not contribute to the growth of the system. A more subtle point is that the inclusion of inter-planetesimal gravity results in collisions that would not otherwise occur in a scenario without gravitational focusing. This results in a number of low speed and accretion enabling collisions.

From these results we find that $e_{ff} > 0.01$ indicates a region of the planetesimal disk that is too perturbed to support planetesimal growth and $e_{ff} < 0.01$ may be calm enough for planetesimal accretion. If we apply this criteria to all known Kepler circumbinary planets, and assuming that this critical value applies to $e_f$ as well, only Kepler 47(AB)c could possibly have formed in-situ and all of the rest: Kepler 34(AB)b, 16(AB)b, 35(AB)b, 38(AB)b, 47(AB)b and 64(AB)b, must have formed further out in the planetesimal disk and migrated inwards to their current location. The effect of a reduced collision rate at large orbital radii on the formation timescale of planets is possibly explained by \citet{alexander12} who suggests that circumbinary planets have more time to form than planets around single stars due to the longer lifetime of disks around short period binaries.

\section{Summary and Further Work}\label{sec:summary}

In this paper we presented results from high-resolution N-body simulations investigating the likelihood of in-situ formation of Kepler-34(AB)b. We began with a unperturbed disk using eccentricity and inclination distributions from numerical simulations of planetesimal evolution around single stars and allowed the system to evolve to quasi-steady state while suppressing collisional evolution of the planetesimals. Eccentricity forcing from the binary pumped up the planetesimal eccentricities to 10 times the initial mean value. After 1000 $P_{AB}$ a quasi-steady state was reached where planetesimals with similar orbital radii take on eccentricity extremes, such that both circularised and eccentric bodies can occupy the same space. This is shown by the speckled effect in Figure \ref{fig:multi}{\bf D}.  This new initial condition was highly perturbed, which meant many orbit crossing events occurred. 

After the initial condition was reached we allowed the planetesimals to evolve collisionally using our new collision model EDACM based on \citep{leinhardt12} which allows erosion, accretion, and bouncing events. The high eccentricity observed in the inner circumbinary disk translates into a dominance of super-catastrophic events seen in Figure \ref{fig:collspatevo}. The impact velocity and parameter decrease with increasing orbital radius; the latter due to a narrower eccentricity dispersion resulting in a reduced number of orbit crossings. The number of high energy, erosive events decrease as a function of increasing semi-major axis. However, planetesimal growth events do not dominate the collision outcomes within the simulation domain ($a<1.5$ AU). We therefore show that the disk is a hostile environment even for our gravitationally strong 120 km planetesimals, suggesting even more difficulties for sustained accretion to occur in simulations that feature much smaller bodies.

Using statistical arguments from collisional data, in addition to physical growth rates, we find that Kepler-34(AB)b would struggle to grow in-situ. In addition, we suggest that from all the known Kepler circumbinary planets only Kepler-47(AB)c could have formed in-situ while the rest must have formed at larger $a$ where the protoplanetary disks were less perturbed by the binary stars and migrated inwards to their current location. We also show that inter-planetesimal gravity must be included in planet formation models in order to capture gravitational focusing effects that may be missed otherwise, such as low-velocity, growth-enabling impacts that may influence the outcome of the simulation.

Previous work which has attempted to hybridise a protoplanetary disk with a gas counterpart has suggested that the gas disk is similarly perturbed by a stellar binary causing further excitations to the planetesimal eccentricities, which, could further provoke unfavourable impact velocities \citep{marzari08,marzari13,paardekooper08}. Our simulations, therefore, likely present a best-case scenario for planetesimal accretion.

\section{Acknowledgements}\label{sec:ack}

The authors acknowledge support from the Science and Technology Facilities Council, The Royal Society and the University of Bristol Advanced Computing Research Centre. 




\end{document}